\documentclass[11pt,nofootinbib]{article}
\pdfoutput=1
\usepackage[english]{babel}
\usepackage{cite,amsmath,amssymb,amsbsy,amstext, amsthm, simplewick}
\usepackage{hyperref}
\usepackage{graphicx}
\usepackage{amsfonts}
\usepackage{amssymb}
\usepackage[small]{caption}
\usepackage{upgreek}
\usepackage{exscale,relsize}
\usepackage[makeroom]{cancel}
\usepackage{soul}
\usepackage{lscape}
\usepackage{indentfirst}
\usepackage{latexsym}
\usepackage{multirow}
\usepackage{tabls}
\usepackage{epsfig}
\usepackage{bm}
\usepackage{mathrsfs}
\usepackage{comment}
\usepackage{xspace}
\usepackage[usenames,dvipsnames]{color}
\allowdisplaybreaks[1]

\allowdisplaybreaks[1]


\numberwithin{equation}{section}
\def\nn{{\nonumber}}
\def\Mp{{M_{\rm Pl}}}
\def\beq{\begin{equation}}
\def\eeq{\end{equation}}

\def\0{{\boldsymbol 0}}
\def\ba{{\boldsymbol{a}}}
\def\bk{{\boldsymbol{k}}}
\def\bq{{\boldsymbol{q}}}
\def\bp{{\boldsymbol{p}}}
\def\bv{{\boldsymbol{v}}}
\def\bx{{\boldsymbol{x}}}

\def\bzero{{\boldsymbol{0}}}

\def\bea{\begin{eqnarray}}
\def\eea{\end{eqnarray}}
\def\pd{ {\partial} }

\DeclareRobustCommand{\SkipTocEntry}[4]{}

\def\addCaltech{\small Theoretical Astrophysics (TAPIR), Walter Burke Institute for Theoretical Physics,\\ California Institute of Technology, Pasadena, California 91125, USA}
\def\addUPitt{\small Pittsburgh Particle Physics Astrophysics and Cosmology Center (PITT PACC)\\Department of Physics and Astronomy, University of Pittsburgh, Pittsburgh, Pennsylvania 15260, USA}
\def\addCMU{\small Department of Physics and Astronomy\\ Carnegie Mellon University, Pittsburgh, Pennsylvania 15213, USA}
\def\addICTP{\small ICTP South American Institute for Fundamental Research,\\ Instituto de F\'isica Te\'orica - Universidade Estadual Paulista\\ Rua Dr. Bento Teobaldo Ferraz 271, 01140-070 S\~ao Paulo, SP Brazil}
\begin{document}

\begin{titlepage}

\setcounter{page}{1} \baselineskip=15.5pt \thispagestyle{empty}

\begin{center}

{\fontsize{18}{26}\selectfont  \sffamily \bfseries 
The tail effect in gravitational radiation-reaction: \\ [-0.2cm]   time non-locality and\\ \vskip 0.4 cm renormalization group evolution}

\end{center}

\vspace{0.1cm}

\begin{center}
{\fontsize{13}{30}\selectfont  Chad R.~Galley,$^{1}$ Adam K. Leibovich,$^{2}$  Rafael A.~Porto$^{3}$ 
and Andreas Ross$^{4}$}
\end{center}

\begin{center}

\vskip 8pt
\textsl{$^1$  \addCaltech}
\vskip 7pt

\textsl{$^ 2$ \addUPitt}
\vskip 7pt
\textsl{$^3$ \addICTP}
\vskip 7pt
\textsl{$^4$ \addCMU}

\end{center}

\vspace{1.2cm}
\hrule \vspace{0.3cm}
\noindent {\sffamily \bfseries Abstract} \\[0.1cm]
We use the effective field theory (EFT) framework to calculate the tail effect in gravitational radiation reaction, which enters at 4PN order in the dynamics of a binary system. The computation entails a subtle interplay between the near (or potential) and far (or radiation) zones. In particular, we find that the tail contribution to the effective action is non-local in time, and features both a dissipative and a `conservative' term. The latter includes a logarithmic ultraviolet (UV) divergence, which we show cancels against an infrared (IR) singularity found in the (conservative) near zone. The origin of this behavior in the long-distance EFT is due to the point-particle limit --shrinking the binary to a point-- which transforms a would-be infrared singularity into an ultraviolet divergence. This is a common occurrence in an EFT approach, which furthermore allows us to use renormalization group (RG) techniques to resum the resulting logarithmic contributions. We then derive the RG evolution for the binding potential and total mass/energy, and find agreement with the results obtained imposing the conservation of the (pseudo) stress-energy tensor in the radiation theory. While the calculation of the leading tail contribution to the effective action involves only one diagram, five are needed for the one-point function. This suggests logarithmic corrections may be easier to incorporate in this fashion. We conclude with a few remarks on the nature of these IR/UV singularities, the (lack of) ambiguities recently discussed in the literature, and the completeness of the analytic Post-Newtonian framework.  
\vskip 10pt
\hrule

 \end{titlepage}
 \tableofcontents

 \section{Introduction}
 
The effective field theory (EFT) framework introduced in \cite{nrgr}, and coined NRGR for `Non-Relativistic General Relativity', has proven to be very successful in the study of the two-body problem in general relativity. Originally, the formalism in \cite{nrgr} was used to derive the first Post-Newtonian correction (1PN) to the conservative dynamics for non-rotating objects. Soon after the 2PN \cite{nrgr2pn} and 3PN \cite{nrgr3pn} gravitational potentials were computed, reproducing previous results within traditional methods, e.g. \cite{3pn1,3pn2,3pn3} (see \cite{Blanchet} for a complete list of references.) On the other hand, in the radiative sector, the 1PN \cite{andirad} and 2PN \cite{andiunp} radiative multipoles were computed within NRGR, which is however still below the state of the art for non-spinning binary systems at 3PN order, e.g. \cite{Blanchet}. NRGR was promptly extended in \cite{nrgrs} to include spin degrees of freedom, and used to describe spinning compact binary systems to 3PN \cite{nrgrs,eih, comment, nrgrss,nrgrs2, nrgrso, srad, amps,Levi:2008nh,Perrodin:2010dy,Levi:2010zu}. Some of these results were previously derived in \cite{owen,buo1,buo2,damournloso} for the spin-orbit sector at 2.5PN order. The spin-spin gravitational potentials to 3PN were obtained within the EFT approach in \cite{eih, comment, nrgrss,nrgrs2,Levi:2008nh}, and in \cite{Schafer3pn,Schafer3pn2,schaferEFT,HergtEFT} and \cite{bohennloss}, using the  Arnowitt-Deser-Misner (ADM) and harmonic gauge formalisms, respectively. The radiative multipole moments quadratic in the spin needed for the radiated power to 3PN, computed in \cite{srad} using the framework of \cite{nrgrs,nrgrs2,andirad,andirad2}, were also obtained in \cite{bohennloss}, although the comparison is pending. The required multipoles for the gravitational wave amplitude to 2.5PN order were computed in \cite{amps}, see also \cite{Buonanno:2012rv}. Higher order effects have been incorporated in the conservative sector. In \cite{levinnlo1,levinnlo2,levinnlo3,equiv4pn} the gravitational spin-orbit and spin-spin potentials were computed at 3.5PN and 4PN order, respectively. These results were derived with more traditional methods in \cite{hartung, bohennloso,steinhoffnnlo1}, except for finite-size effects \cite{levinnlo3}, which are more efficiently handled in an EFT framework \cite{nrgr, nrgrs,nrgrs2}. The~formalism in \cite{nrgrs,nrgrs2} was also used to compute the leading finite size effects cubic (and quartic) in the spin in \cite{eftvaidya,levis3,marsats3}. In conjunction, all of these results augment the knowledge of the dynamics of binary compact objects to 4PN order. For thorough reviews on the two-body problem and the EFT approach see \cite{Blanchet,Buoreview,ALT,nrgrLH,riccardocqg,portoreview,chadreview}.\vskip 4pt  

The computation of the (local part of the) spin independent 4PN potential, at next-to-next-to-next-to-next-to leading order beyond the Newtonian approximation, was recently culminated in \cite{4pn,4pndim,4pn1,4pn2} and \cite{4pnluc} using the ADM and harmonic formalisms, respectively. A partial result computed in NRGR~\cite{nrgr4pn} has shown full agreement. However, the subtleties associated with infrared (IR) and ultraviolet (UV) divergences, which appear at this order, have led to a disagreement between different approaches \cite{4pnluc,4pnDam2}. As we shall see, the present paper partially addresses some of these issues --in particular the (lack of) ambiguities and completeness of the PN framework-- pending the completion of the full 4PN conservative dynamics within NRGR. At~4PN order there is also a contribution to the effective action which is non-local in time, e.g. \cite{Blanchet,tail3n}, recently revisited in \cite{4pn2,4pnluc}, as well as logarithmic corrections to the binding mass/energy. The latter were obtained within NRGR in \cite{andirad3} through the conservation of the (pseudo) stress-energy tensor in the radiation zone, and in full agreement with a previous computation in \cite{ALTlogx}. Both of these results feature prominently in this work, but instead arise from the computation of radiation-reaction effects. \vskip 4pt 

The study of time-irreversible back-reaction effects within the EFT formalism was initiated in \cite{chadgsf} in the extreme mass ratio limit, and in\cite{chadbr1} for NRGR, by implementing the classical limit of the `in-in' formalism, e.g. \cite{inin1, inin2}. Later, the radiation-reaction force to 3.5PN order \cite{Iyer1,Iyer2} was rederived within the EFT approach in \cite{chadbr2} using a framework that extends Hamilton's principle to generic nonconservative systems \cite{chadprl, chadprl2}. These results are obtained at leading order in $G_N$ in the radiation theory. In the present work we incorporate non-linear effects in the radiation zone by computing the tail contribution, e.g. \cite{tail1,tail2,tail3,tail4,tail5}, to the effective action. We will find that the tail contribution plays an essential role in both the results mentioned above, namely, the presence of logarithms and time non-locality. \vskip 4pt 

The non-linear couplings due to the higher order tails in gravitational wave emission produce divergences. The IR singularities (which are also present in the leading tail contribution) exponentiate into an overall phase in the amplitude \cite{andirad}, which drops out of the total radiated power or can be removed from the gravitational waveform via a time redefinition \cite{amps}. On the other hand, the UV divergences thus far have been properly renormalized through counter-terms in the radiation theory, which led to renormalization group (RG) trajectories for the binding mass/energy and multipole moments, described in \cite{andirad,andirad3}. As~we discuss here, similar behavior unfolds through the study of radiation-reaction effects, albeit involving a subtle interplay between the theory of potential modes (near zone) and the radiation sector (far zone).  \vskip 4pt

The radiation-reaction force corrects the dynamics of the constituents of the binary system at the orbit scale, $r$. However, we compute it by integrating out the radiation field, $\bar h_{\mu\nu}$, with a long-distance effective action \cite{andirad,andirad2} 
\bea
\label{eq:lag}
 S_{\rm eff}^{\rm rad}[\bx_a, \bar h_{\mu\nu}]  =  -\int dt \sqrt{\bar g_{00}} && \hspace{-0.5cm}\bigg[M(t)-  \sum_{\ell=2} \bigg( \frac{1}{\ell!} I^L(t) \nabla_{L-2} E_{i_{\ell-1}i_\ell}\\ &&+\frac{2\ell}{(2\ell+1)!}J^L(t) \nabla_{L-2} B_{i_{\ell-1}i_\ell}\bigg)\bigg]\,,\nonumber
\eea
where radiation modes vary on scales of order $\lambda_{\rm rad} \sim r/v \gg r$ and propagate on a background (Schwarzschild) geometry sourced by the first term, $M$, which at leading order gives a potential,
\beq
\label{Phist}
\Phi (\bq)\simeq -\frac{G_N M}{\bq^2}\,.
\eeq
Therefore, the study of radiation reaction entails the interaction between different zones. This is even more relevant when the tail contribution is incorporated, as we show here.\vskip 4pt

After integrating out the radiation field, including the tail effect, we will find that the resulting effective action for the dynamics of the binary is non-local in time, in agreement with a recent claim in \cite{4pn2}. We also find both dissipative and `conservative' contributions, and the latter includes the presence of a logarithmic UV divergence. Hence, unlike the renormalization of the one-point function in \cite{andirad,andirad3} which occurs in the radiation zone, the counter-term for this divergence must originate in the potential region. We argue this involves 
the existence of an IR singularity in the near zone. This is expected because the radiation-reaction potential is now part of the dynamics at short(er) distances. Moreover, UV divergences in the theory of potentials are removed by counter-terms in the worldline theory for each constituent in the binary and, because of the `effacement theorem,' do not contribute until 5PN order (for non-rotating bodies). \vskip 4pt The seed of the UV divergence in the tail computation is the point-particle limit, implicitly taken in \eqref{eq:lag}, where the binary as a whole is treated as a point-like source. By shrinking the binary to a point we transform a would-be IR singularity into a UV divergence. This is a common occurrence in an EFT approach that, furthermore, allows us to use RG techniques to resum the resulting logarithmic contributions. We then derive the RG evolution for the binding potential and total mass/energy and find agreement with the results obtained in \cite{andirad3}. While the calculation of the radiation-reaction potential involves computing only one diagram, five are needed for the one-point function in \cite{andirad3}, which suggests higher order logarithmic terms may be easier to incorporate in this fashion.\vskip 4pt 

This paper is organized as follows. We first review the computation of radiation-reaction effects within NRGR. We then integrate out the radiation field including the tail effect, and demonstrate the presence of dissipative and conservative terms, and the time non-locality of the effective action. Afterwards we discuss renormalization and RG equations. We conclude with a few remarks on the breakdown of the separation of scales, the origin of the ambiguities recently discussed in the literature, e.g. \cite{4pnDam2}, and the completeness of the analytic PN framework. The computation of the tail effect in the radiation-reaction potential within the EFT formalism was first approached in \cite{tailFoffa}. We also comment at the end on the main differences between \cite{tailFoffa} and the present work. We~relegate details of the computation to an appendix.

\section{Gravitational radiation-reaction in NRGR}
\label{sec:rr}

Accommodating the time-asymmetric interactions associated with nonconservative processes, like radiation reaction, at the level of the action entails formally doubling the degrees of freedom in the problem so that $\bx_a \to \{\bx^{(1)}_a,\bx^{(2)}_a\}$, with $\bx_a$ being the physical coordinates of the $a^{\rm th}$ body, and similarly for the radiative metric perturbations, $\bar{h}_{\mu\nu} \to \{ \bar{h}_{\mu\nu}^{(1)}, \bar{h}_{\mu\nu}^{(2)} \}$. 
After the latter are integrated out from the theory, we will be left with an effective action that can be written as
\begin{align}
	W [ \bx^\pm_a ] = \int dt \, \big( L [\bx_a^{(1)} ] - L [\bx_a^{(2)} ] + R [ \bx_a^{(1)}, \bx_a^{(2)} ] \big),
\end{align}
where $L = \int dt (K - V)$ is the usual Lagrangian that accounts for the binary's conservative interactions while $R$ accommodates non-conservative effects, such as radiation reaction. It is worth noticing that if $R$ contains terms that can be written in a manner resembling the first two terms, namely, 
\begin{align}
	R [ \bx_a^{(1)}, \bx_a^{(2)} ] \supset F[\bx_a^{(1)}] - F[\bx_a^{(2)}],
\label{eq:addsep}
\end{align}
then $F$ may be absorbed into a redefinition of $L$ and, ultimately, the conservative binding potential~\cite{chadprl}.
This observation will be important later on when we discuss the tail effect in Sec.~\ref{sec:tail}.
Details of the underlying theory of general nonconservative mechanics is given in~\cite{chadprl} and extended to field theories and continuum systems (including viscous fluid flows with entropy production) in \cite{chadprl2}. \vskip 4pt

The leading contribution to the radiation reaction force comes from the following diagram in the effective action \cite{chadbr1}, 
\vspace{-0.4cm}
\begin{align}
\begin{aligned}
\label{eq:br2}
iW[\bx^\pm_a]   & =  \parbox{21mm}{\includegraphics[width=0.2\textwidth]{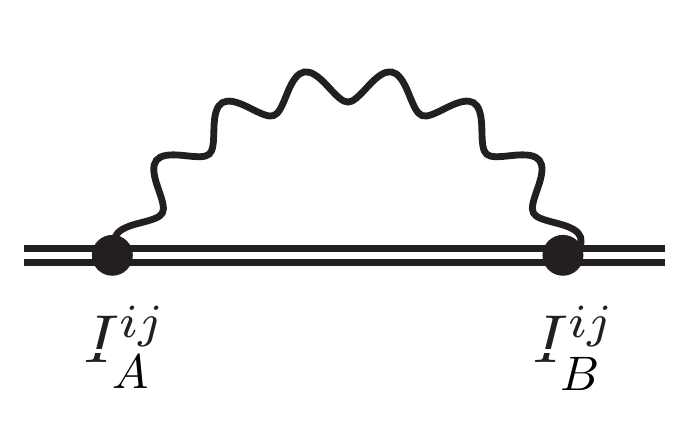}}  \\
& =  \left( \frac{1}{2} \right) \left( \frac{i}{2 \Mp} \right)^2 \int dt \int dt^\prime I_A^{ij}(t) \big\langle E^A_{ij} (t, \bzero) E^B_{kl}(t^\prime, \bzero) \big\rangle I_B^{kl}(t^\prime),
\end{aligned}
\end{align}
where $A, B = \pm$, $\bx_{a +} \equiv ( \bx_a^{(1)} + \bx_a^{(2) } ) /  2$, and $\bx_{a -} \equiv \bx_a^{(1) } - \bx_a^{(2) }$. The tensor $E_{ij} (t, \bzero)$ is the electric part of the Weyl curvature tensor evaluated at the binary's center of mass, which is taken to be at the origin.
The equations of motion are found from the effective action through
\begin{align}
	\left[ \frac{ \delta W }{ \delta \bx^i_{a-} (t) } \right]_{\rm PL} = 0  \qquad \Longrightarrow \qquad \frac{d}{dt} \frac{ \pd L }{ \pd \bx^i _a } - \frac{ \pd L }{ \pd \bx^i _a } = \bigg[ \frac{ \pd R }{ \pd \bx^i_{a-} } - \frac{ d }{dt} \frac{ \pd R }{ \pd \bv^i_{a-} } \bigg]_{\rm PL} \, ,
\label{eq:actin}
\end{align}
where ``PL'' indicates the physical limit wherein $\bx_{a-} \to 0$ and $\bx_{a+} \to \bx_a$. The two-point function in the harmonic gauge for trace-reversed metric perturbations is given by\begin{align}
	\big\langle E^A_{ij} (t, \bzero) E^B_{kl}(t^\prime, \bzero) \big\rangle = -\frac{i}{8} \bigg[ &  \pd_i \pd_j \pd_{k'} \pd_{l'} + 2 P_{ij k l} \pd_0^2 \pd_{0'}^2 - \eta_{ij} \pd_0^2 \pd_{k'} \pd_{l'} - \eta_{kl} \pd_i \pd_j \pd_{0'}^2  +   \eta_{ik} \pd_j \pd_0 \pd_{l'} \pd_{0'} \nn \\
		& + \eta_{jl} \pd_i \pd_0 \pd_{k'} \pd_{0'} + \eta_{il} \pd_j  \pd_0 \pd_{k'} \pd_{0'} + \eta_{jk} \pd_i \pd_0 \pd_{l'} \pd_{0'}  \bigg] G^{AB} (t-t', \bzero),
\end{align}
where  the prime on a spacetime index of a derivative is taken with respect to $x'{}^\mu$, and
\begin{align}
	P_{\alpha \beta \gamma \delta} = \frac{1}{2} \big( \eta_{\alpha \gamma} \eta_{\beta \delta} + \eta_{\alpha \delta } \eta_{\beta \gamma} - \eta _{\alpha \beta} \eta_{\gamma \delta} \big).
\end{align}
The matrix of propagators\footnote{The propagator's tensorial structure factors into $P_{\alpha \beta \gamma \delta}$ and a scalar Green's function in this gauge.} in the $\pm$ variables is
\begin{align}
\label{eq:GreenAB}
	G^{AB} (x^\mu - x'{}^\mu ) = \left( 
		\begin{array}{cc} 
			0 & G_{\rm adv} (x^\mu - x' {}^\mu) \\
			G_{\rm ret} (x^\mu - x'{}^\mu ) & 0
		\end{array} 
	\right),
\end{align}
with $G^{-+}(x^\mu-x'{}^\mu) = G_{\rm ret}(x^\mu-x'{}^\mu)$. The two propagators are needed to enforce the causal (i.e., outgoing) boundary conditions on the metric perturbations being integrated out \cite{chadprl}.\vskip 4pt

Computing the diagram in \eqref{eq:br2} results in \cite{chadbr1}
\begin{align}
	W [\bx_a^\pm]= - \frac{ G_N }{ 5 } \int dt \, I_{-}^{ij} (t) I_{+ij} ^{(5)} (t)   \equiv \int dt \, R_{\rm rad} [ \bx_a^\pm],
\label{eq:WBT}
\end{align}
where the superscript $(5)$ indicates five time derivatives and we introduced 
\begin{align}
	I_{-}^{ij} (t) & \equiv I^{ij} (t; \bx_a^{(1)}) - I^{ij} (t; \bx_a^{(2)} )  =  \sum_a m_a \left(\bx_{a-}^{i} \bx_{a+}^{j} +  \bx_{a+}^{i} \bx_{a-}^{j} - \frac{2}{3}\, \delta^{ij} \bx_{a-} \! \cdot \! \bx_{a+}\right)  + {\cal O} (\bx_{a -}^3 ) , \nn \\
	I_{+}^{ij} (t) & \equiv \frac{1}{2} \Big( I^{ij} (t; \bx_a^{(1)}) + I^{ij} (t; \bx_a^{(2)} ) \Big) = \sum_a m_a \left( \bx_{a+}^i \bx_{a+}^j - \frac{1}{3} \delta^{ij} \bx_{a+}^2 \right) + {\cal O} (\bx_{a -}^2 ).
\end{align}
Using \eqref{eq:actin} we obtain the acceleration on the $a^{\rm th}$ body resulting from \eqref{eq:WBT} as \cite{chadbr1}
\begin{align}
\label{eq:arad}
	(\ba^i_a)_{\rm rr}(t) = - \frac{2 G_N}{5} I^{(5)ij}(t) \bx_a^j(t) .
\end{align}
This is precisely the radiation-reaction force derived by Burke and Thorne \cite{thorneBT1,thorneBT2}. At this order, notice that $R_{\rm rad}$ defined in (\ref{eq:WBT}) cannot be absorbed into a redefinition of the binding potential and thus represents a truly non-conservative effect.   \vskip 4pt
The action for the conservative sector of the theory (i.e., `turning off' radiative effects) is invariant under time translations implying the existence of a {\it conserved} quantity, namely, the binary's binding mass/energy,\footnote{More generally, higher order time derivatives, e..g. accelerations, may be present in the effective action. If~ these are not reduced using lower order equations of motion, the expression for the binding mass/energy in \eqref{eq:Noether0} has to be modified accordingly, see e.g. \cite{3pnEL}.}
\begin{align}
	M \equiv \sum_a \bv_a \cdot \frac{ \pd L }{ \pd  \bv_a } - L \, .
\label{eq:Noether0}
\end{align}
Once radiation is turned on, $M$ is no longer conserved since $R[\bx_a^\pm]$ accounts for time-irreversible interactions\cite{chadprl2, chadreview}. Hence, we have 
\begin{align}
	\dot{M} = \sum_a \bv_a \cdot  \bigg[ \frac{ \pd R }{ \pd \bx_{a-} } - \frac{ d }{dt } \frac{ \pd R }{ \pd \bv_{a-} }  \bigg]_{\rm PL} .
\label{eq:Mdot13}
\end{align}
For the case of gravitational radiation reaction, using $R_{\rm rad}$ from \eqref{eq:WBT} we find
\begin{align}
	\dot{M} = - \frac{ G_N }{ 5 } I^{(1)ij}(t) I^{(5) ij} (t)\,,
\label{eq:powerLOSF}
\end{align}
at leading PN order. Notice, after some simple algebraic manipulations, we can write~\eqref{eq:powerLOSF} as 
\beq
\dot E = - \frac{G_N }{ 5}   {I}^{(3)ij} (t) I^{(3)ij}(t) \,,
\eeq
where
\begin{align}
	E = M + \frac{G_N}{5}  I^{(1)ij} (t) I^{(4)ij}(t) - \frac{G_N}{5}  {I}^{(2)ij} (t) I^{(3)ij}(t)\,.
\end{align}
The extra pieces are analogous to the Schott energy in electrodynamics, and may be interpreted in terms of near-zone contributions from the metric perturbations \cite{chadreview}.\footnote{The expressions for $E$ and $\dot{E}$ can be derived directly from the effective action, see \cite{chadreview}. This is beyond the scope of this paper.}\vskip 4pt

We then average \eqref{eq:powerLOSF} over a bound (not necessarily circular) orbit and find that the Schott-like terms average away leaving behind the well-known quadrupole formula,
\begin{align}
\label{eq:powerLO}
	\big\langle \dot{M} \big\rangle = - P_{LO}  = - \frac{G_N }{ 5} \big\langle  {I}^{(3)ij} (t) I^{(3)ij}(t) \big\rangle\, .
\end{align}
The above result was also derived through the conservation of the (pseudo) stress-energy tensor in~\cite{andirad3}.\vskip 4pt

The previous steps can be generalized to all $\ell$-order radiative multipoles. The diagrams contributing to the effective action are
\beq
\hspace{-5cm} 
	iW[\bx_a^{(\pm)}] \, = \,  \sum_{\ell \ge 2} ~~ \parbox{21mm}{\includegraphics[width=0.45\textwidth]{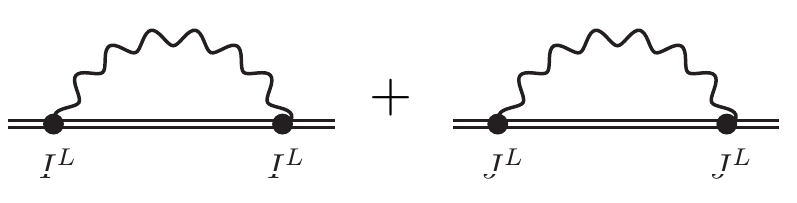}} \nn
\eeq
and are found to give \begin{align}
\begin{aligned}
\label{eq:Wtailall}
	W[\bx_a^{\pm}] = G_N \sum_{\ell \ge 2} \frac{(-1)^{\ell+1}(\ell+2)}{(\ell-1)}  \int dt \, \bigg( & \frac{2^\ell (\ell + 1)}{\ell (2 \ell + 1)!} I_{-}^L(t) I_{+}^{L \, (2\ell + 1)}(t)  \\ 
		&+  \frac{2^{\ell+3} \ell}{(2 \ell + 2)!}  J_{-}^L(t)  J_{+}^{L \, (2\ell + 1)}(t)\bigg) \,,
\end{aligned}
\end{align} 
which incorporates back-reaction effects at leading order in $G_N$ in the far zone. From \eqref{eq:Wtailall} one can derive any quantity of interest in the radiation region at linear order in $G_N$, such as the corresponding radiation-reaction forces and orbit-averaged balance equations for energy and angular momentum for compact binary inspirals.

\section{The tail effect}
\label{sec:tail}
 
We now move on to incorporating non-linear gravitational interactions in the radiation zone, and the contribution from the tail effect to the radiation-reaction potential. The relevant Feynman diagram is shown in Fig.~\ref{nltail}. We first demonstrate the time non-locality, together with the existence of dissipative and conservative contributions, to the effective action. Subsequently we discuss the renormalization and RG evolution equations.

\subsection{Time non-locality}

\begin{figure}[t!]
\centerline{{\includegraphics[width=0.3\textwidth]{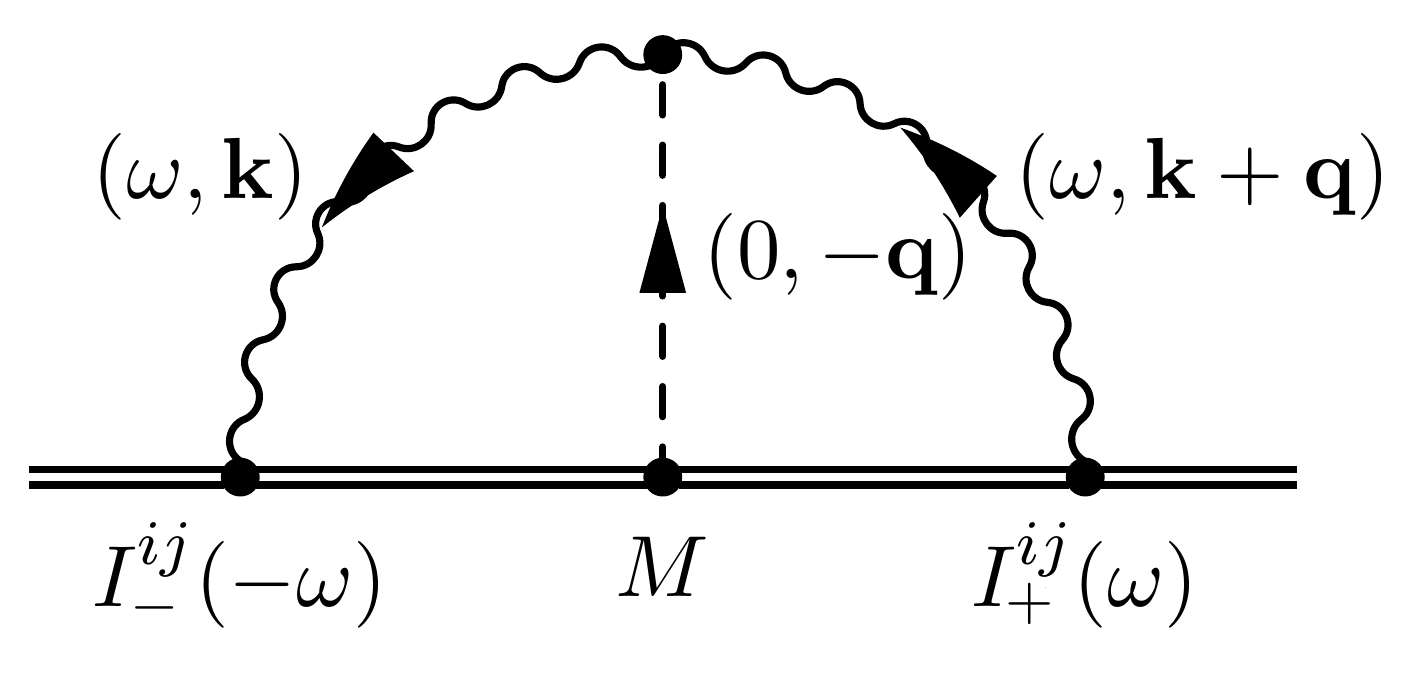}}}
\caption[1]{Feynman diagram for the tail contribution to the radiaction-reaction force. The $M$ in the tail correction may be taken as the leading averaged binding mass/energy, since the scale (and time) dependence enters at higher orders.} \label{nltail}
\end{figure}

The resulting two-loop integral(s) arising from Fig.~\ref{nltail} can be written schematically as
\beq
 i W_{\rm tail}  =  \int \frac{d\omega}{2\pi} \int_{\bk, \bq} M I^{ij}_-(- \omega) I^{ij}_+(\omega) V_{\bar h\bar h\Phi}(\omega, \bk, \bq) \frac{i}{- \bq^2 } \frac{i}{(\omega + i \epsilon)^2- \bk^2} \frac{i}{(\omega + i \epsilon)^2- (\bk + \bq)^2} \, . \label{eq:tailRR}
\eeq
Here, $V_{\bar h \bar h\Phi}$ represents the three-graviton coupling, and $\int_\bp \equiv \int \frac{d^3\bp}{(2\pi)^3}$. Notice the retarded boundary conditions in the pole structure of the propagators. We use dimensional regularization (dim.~reg.) and, after some laborious manipulations outlined in App.~\ref{sec:app1}, we arrive at 
\begin{align}
 i W_{\rm tail}[\bx_a^\pm] = -i \int \frac{d\omega}{2 \pi} \,  \frac{(d-3) M \omega^4 \, I_-^{ij}(-\omega) I_+^{ij}(\omega)}{32 (d-2)^2 (d-1)(d+1)} \Big[(d^2 - 2 d + 3) I_0 - \frac{d(d-2)(d-1)}{d-4} \omega^2 J_0 \Big]\,,
\end{align}
in terms of two $d$-dimensional integrals, $I_0$ and $J_0$, see \eqref{eq:I0}-\eqref{eq:J0}. The result is UV divergent. Expanding around $d=4$ we find,\footnote{The renormalization scale $\mu$ in the logarithms appears from the shift in the mass-dimension of the couplings in the theory, in our case $G_N$, in $d$ spacetime dimensions. See \cite{andirad} for more details.} 
\begin{align}
W_{\rm tail}[\bx_a^\pm]  =   \, \frac{2 G_N^2 M}{5} \int_{-\infty}^\infty \frac{d\omega}{2 \pi} \,  \omega^6 \, I_-^{ij}(-\omega) I_+^{ij}(\omega) \bigg[ & -\frac{1}{ (d-4)_{\rm UV}} - \gamma_E + \log\pi \nn\\ 
  	&-  \log \frac{\omega^2}{\mu^2} + \frac{41}{30} + i \pi \, \text{sign}(\omega) \bigg] .
\label{eq:RRnl}
\end{align}

The pole is removed by a  counter-term (see below) and we obtain
 \begin{align}
W_{\rm tail}[\bx_a^\pm] & = - \frac{2 G_N^2 M}{5}\int \frac{d\omega}{2 \pi} \,  \omega^6 \, I_-^{ij}(-\omega) I_+^{ij}(\omega) \left[\log \frac{\omega^2}{\mu^2} - i \pi \,\text{sign}(\omega) \right] \, ,\label{eq:WtailRR}
\end{align}
where we also absorbed a constant piece into a redefinition of $\mu$.
We can now Fourier transform from frequency space back to the time domain, yielding   
\begin{align}
W_{\rm tail}[\bx_a^\pm] = \frac{4 G_N^2 M}{5} \Big(&{\rm PV}\int dt \, I_-^{(3)ij}(t) \int_{-\infty}^t dt' \, I_+^{ij(3)}(t') \left[\frac1{t-t'} \right]  \label{finitetail} \\ &+  \, \int dt \, I_-^{(3)ij}(t) I_+^{ij(3)}(t) \log \mu\Big) \equiv \int dt\, R_{\rm tail}[\bx_a^\pm]\nn\,,
\end{align}
where `PV' stands for Principal Value. We can also write the effective action as,
\beq
\label{finitetail2}
W_{\rm tail}[\bx^{\pm}_a] = \frac{4 G_N^2 M}{5}~{\rm PV} \int dt  \, I_-^{(3)ij}(t) \int_{-\infty}^t dt' \, I_+^{ij(4)}(t')\log(|{t-t'}|\mu)\,.\eeq
This result is formally equivalent to the non-local term discussed in \cite{4pn2,4pnDam2} and \cite{4pnluc} (see also \cite{tail3n}).\vskip 4pt  

\subsection{Conservative and dissipative terms}

It is easy to see that all the terms in \eqref{eq:RRnl} are of the form in~\eqref{eq:addsep} except for the one involving $\text{sign}(\omega)$. (This is clear since it is the only term that is not invariant under $\omega \to -\omega$.) Therefore, $W_{\rm tail}$ contains both conservative and dissipative interactions. In particular, the pole and logarithm terms are both part of the conservative sector and consequently renormalize the binary's binding mass/energy. This implies that the RG structure of the theory, which we discuss in the next sub-section, occurs entirely in the conservative sector and the dissipative term is finite.\vskip 4pt

From the expressions in \eqref{finitetail} (or \eqref{finitetail2}) and \eqref{eq:actin}, we then derive the contribution due to the conservative  and non-conservative terms to the radiation-reaction acceleration,
\beq
\left(\ba^j_{a}\right)_{\rm rr,tail} (t)= \left(\ba^j_{a}\right)_{{\rm cons}} (t,\mu) + \left(\ba^j_{a}\right)_{{\rm diss}} (t)\,,
\eeq
where 
\bea
 \left(\ba^j_{a}\right)_{{\rm cons}} (t,\mu) &=& -\frac{4 G_N^2 M}{5} \, \bx_a^i(t)\left( I^{ij(6)}(t)\log \mu^2 + {\rm PV} \int_{-\infty}^\infty dt' \, I^{ij(6)}(t') \left[\frac{1}{|t-t'|}\right]\right)\,,\label{eq:tailac}\\
 \left(\ba^j_{a}\right)_{{\rm diss}} (t) &=& -\frac{4 G_N^2 M}{5} \, \bx_a^i(t) ~{\rm PV} \int_{-\infty}^\infty dt' \, I^{ij(6)}(t') \left[\frac{1}{t-t'}\right]\,.
\label{eq:tailanc}
\eea
 Notice both combine to a causality-preserving force, as expected. In other words, the presence of both conservative and non-conservative terms guarantees the integral in  \eqref{finitetail} (and \eqref{finitetail2}) only receives contributions from $t' < t$.

\subsection{Renormalization}

\subsubsection{Counter-term}

After using dim.~reg., the computation of the tail effect in gravitational radiation reaction contains a UV pole. Therefore, we require a counter-term to remove the divergence when $d\to 4$. 
Since the pole appears in the conservative sector (see above) we require the following counter-term
\begin{equation}
 -\int dt \, V_{\rm ct}[\bx_a] = \, \, \frac{1}{(d-4)_{\rm UV}} \frac{G_N^2 M}{5} \int dt \, I^{(3)ij}(t) I^{(3)ij}(t)\,.\label{eq:Vpmct2}
\end{equation}
(Note the expression in \eqref{eq:Vpmct2} is half of the one in \eqref{eq:RRnl}. This occurs after translating from the minus to the standard variables.)\vskip 4pt
 
The origin of this counter-term, however, is subtle. That is because the divergence in \eqref{eq:Vpmct2} cannot be associated with short-distance behavior in the theory of potentials, which are instead responsible for finite size effects for extended objects. Moreover, the leading order finite size effects for (non-rotating) binary systems enters at 5PN, e.g. \cite{chadreview,portoreview}, whereas \eqref{eq:Vpmct2} contributes at 4PN order. Nevertheless, the UV divergence in \eqref{eq:RRnl} arises in a point-particle limit, the one in which we shrunk the binary to a point-like source, by sending the separation between constituents to zero (represented by the double line in Fig.~\ref{nltail}). However, the separation is kept finite at the orbital scale since it is the typical scale of variation of the potential modes. For the latter, modes in the radiation zone are soft(er). Therefore, it is natural to expect the UV divergence in \eqref{eq:RRnl} to be related to an IR singularity at the orbital scale. Indeed, the existence of such an IR divergence in the theory of potentials was recently found in \cite{4pn1,4pn2,4pnluc}, in both the ADM and harmonic frameworks. The resulting potential, $V_{\rm 4pn}$, may be then split into a local term and IR-dependent pieces \cite{4pn1,4pn2} 
\bea
\label{eq:V4pn1}
V_{\rm 4pn}[\bx_a]  &=&  V_{\rm 4pn}[\bx_a,\tilde\mu] - \frac{1}{(d-4)_{\rm IR}}\frac{G_N^2 M}{5} I^{(3)ij}(t) I^{(3)ij}(t) \,,
\eea
using dim. reg., with
\bea
V_{\rm 4pn}[\bx_a,\tilde\mu] &=& V_{\rm 4pn}^{\rm local}[\bx_a] + \frac{2G_N^2 M}{5} I^{(3)ij}(t) I^{(3)ij}(t)  \log \tilde\mu r \,, \label{eq:V4pn2}
\eea
where $r \equiv |\bx_1-\bx_2|$, and up to a rescaling of $\tilde\mu$ to absorb some extra constants.\,\footnote{Notice there is a factor of $2$ and a relative sign difference between the coefficient of the IR pole and the coefficient of the logarithm. This is often the case in dim.~reg., see e.g.~\eqref{eq:RRnl}. This can also be directly seen in the regularization procedure described in \cite{4pn1,4pndim}, see Eqs.~(A44)-(A53) of \cite{4pn1}.} While the form of $V_{\rm 4pn}^{\rm local}[\bx_a]$ depends on the choice of gauge, the coefficient of the logarithm is physical (and gauge invariant to this order) since, as we shall see, it contributes to the total binding/mass energy of the binary system. Therefore, as we see in \eqref{eq:V4pn1}, the computations in \cite{4pn1,4pn2,4pndim,4pnluc} provide the counter-term needed to cancel the divergence in 
$W_{\rm tail}$.
We have distinguished $\tilde\mu$ from $\mu$ to emphasize the arbitrariness of the renormalization procedure and the choice of subtraction scale, both at the orbital and radiation zones. We return to this issue in sec.~\ref{sec:disc}.

\subsubsection{Renormalization group flow}

After the divergences are subtracted away, the effective action becomes a function of a renormalized Lagrangian, and is given  (in frequency space) by
\begin{align}
	W[\bx_a^\pm]  = \int \frac{d\omega}{2\pi} \Bigg( & L_{\rm ren}[\bx_a^{(1)} ; \omega,\mu] - L_{\rm ren} [ \bx_a^{(2)}; \omega, \mu]  \nn \\
		& - \frac{2 G_N^2 M}{5} \omega^6 \, I_-^{ij}(-\omega) I_+^{ij}(\omega) \left[\log \frac{\omega^2}{\mu^2} - i \pi \text{sign}(\omega) \right]\Bigg), 
\label{eq:WtailRR2}
\end{align}
where
\beq 
	L_{\rm ren} [\bx_a; \omega,\mu]  \equiv K [\bx_a ; \omega]  - V_{\rm ren}[\bx_a ; \omega,\mu] \,,
\label{eq:Lren} 
\eeq 
after including the binary's kinetic term, $K$.
We can then read off the RG evolution equation from the $\mu$-independence of the effective action, 
\beq
\label{eq:Vrenmu}
\mu \frac{\partial}{\partial \mu}W[\bx_a^\pm] = 0 ~~ \Longrightarrow ~~  \mu \frac{\partial}{\partial \mu} V_{\rm ren}[\bx^\pm_a;\omega,\mu] = \frac{4 G_N^2 M}{5}\omega^6 I_-^{ij}(-\omega) I_+^{ij}(\omega)\,,
\eeq
In terms of the standard variables and Fourier transforming back to the time domain, we find the equivalent expression
\beq
\label{eqVtm0}
	\mu \frac{\partial}{\partial \mu} V_{\rm ren}[\bx_a;t,\mu] = \frac{2 G_N^2 M}{5} I^{(3)ij}(t) I^{(3)ij}(t)\,.
\eeq

We may consider, for instance, the case of circular orbits. Then, choosing $\mu \simeq \lambda_{\rm rad}^{-1}$ together with $\mu_0 \simeq r^{-1}$ for the matching scale, we find (using $\mu/\mu_0 \simeq v$)
\beq
\label{eqVtm}
V_{\rm ren}[\bx_a; t,\mu] = V_{\rm ren} [\bx_a; t,\mu_0] +  \frac{2G_N^2 M}{5} I^{(3)}_{ij}(t) I^{(3)}_{ij}(t) \log v\, .
\eeq
This expression is in accordance with the results in \cite{4pn2}. The renormalized potential at $\mu_0 \simeq 1/r$ must be obtained by matching at the orbital scale. We may proceed as follows. First, notice that by choosing $\tilde\mu \simeq 1/r$ in \eqref{eq:V4pn2} we remove the logarithmic contribution. Then, after matching, we get 
\beq
\label{eq:nontrV}
V_{\rm ren} [\bx_a;t,\mu_0\sim 1/r] =  V_{\rm 4pn}^{\rm local}[\bx_a] +  C \, \frac{2G_N^2 M}{5}\, I^{(3)ij}(t) I^{(3)ij}(t)\,.
\eeq
The factor of  $C\equiv \log (\tilde \mu/\mu_0) \sim 1$ accounts for the arbitrariness in the choice of renormalization schemes. The value of $C$ may be obtained, for instance, by comparison with a numerical computation or (semi-) analytically through the self-force program, e.g. \cite{4pn2,binigsf}. For example, according to \cite{4pn2} one finds $C_{\rm ADM} =- \tfrac{1681}{1536}$ in the ADM formalism. A similar constant, $\alpha = \tfrac{811}{672}$, appears in the harmonic framework \cite{4pnluc}.\footnote{While the coefficient of the logarithmic is physical and gauge invariant to this order, the local term in \eqref{eq:nontrV} depends on the choice of gauge. Therefore, in the (background) harmonic gauge which we use here, the resulting value for the constant $C$ in our case may differ from both the values discussed in \cite{4pn2,4pnluc}.}  The existence of this arbitrariness signals the breakdown of the separation of scales between potential and radiation regions. However, this breakdown does not necessarily mean additional information is needed, as advocated in \cite{4pnDam2}. On~the~contrary, it is instead a signature of `double-counting.' Once this is properly addressed, no extra matching condition is necessary. We add a few extra remarks in sec.~\ref{sec:disc}, and will elaborate further on this point elsewhere \cite{zerobinNRGR}.\vskip 4pt  Despite this fact, we can still use the EFT computation to extract information about the dynamics, including logarithmic contributions to the binding mass/energy, which are universal.  Concentrating on the conservative piece, and using \eqref{eq:Mdot13} together with \eqref{eq:tailac}, we may write an energy balance equation 
\beq
\label{eq:balance}
\dot M_{\rm ren}(t,\mu)= \sum_a m_a  \left(\ba_{a}\right)_{{\rm cons} }\cdot \bv_a +\cdots  =   \frac{2 G_N^2 M}{5} \, I^{ij(1)}(t) \int \frac{d\omega}{2\pi} I^{ij(6)}(\omega) e^{i\omega t} \log \frac{\omega^2}{\mu^2}+\cdots\,,
\eeq 
where the renormalized binding mass/energy is given by, see \eqref{eq:Noether0}, 
\beq 
	M_{\rm ren}(t, \mu) \equiv \sum_a \bv_a \cdot \frac{ \pd}{ \pd \bv_a }L_{\rm ren}[\bx_a;t,\mu]  - L_{\rm ren}[\bx_a;t,\mu]  \, .
\eeq   
The ellipsis in \eqref{eq:balance} include also other (non-conservative) terms responsible for the power loss on gravitational wave emission, as we discussed in the previous section. We then return to the case of circular orbits with angular frequency $\Omega$ and take a time average. As it was discussed in \cite{andirad}, the multipole moments have support at the typical scale of gravitational wave radiation, $\lambda_{\rm rad}^{-1} \simeq 2\Omega$, so that $I^{ij}(\omega) \propto \delta(\omega \pm 2\Omega)$. Hence, \eqref{eq:balance} becomes 
\beq
\label{eq:balance2}
\left\langle\dot M_{\rm ren}(t,\mu)\right\rangle =  -\frac{4 G_N^2 M}{5} \,\left\langle I^{ij(1)}(t) I^{ij(6)}(t)\right\rangle\, \log\, (\lambda_{\rm rad} \mu)+\cdots\,.
\eeq
From here, using
\beq
 I^{(1)}_{ij}(t) I^{(6)}_{ij}(t) = \frac{d}{dt} \left( I^{(5)}_{ij}(t) I^{(1)}_{ij}(t) -  I^{(4)}_{ij}(t) I^{(2)}_{ij} (t)+ \tfrac{1}{2} I^{(3)}_{ij}(t) I^{(3)}_{ij}(t)\right)\,, 
 \eeq
on the right-hand side of \eqref{eq:balance}, we get for the {\it conservative} binding energy,
\beq
\label{eq:Eb}
E \equiv M_{\rm ren}(t,\mu) + \frac{2G_N^2 M}{5} \left(2I^{(5)}_{ij}(t) I^{(1)}_{ij}(t) -  2I^{(4)}_{ij}(t) I^{(2)}_{ij} (t)+ I^{(3)}_{ij}(t) I^{(3)}_{ij}(t)\right) \, \log\, (\lambda_{\rm rad} \mu)\,.
\eeq
Notice, at the radiation scale we have $E = M_{\rm ren}(t,\mu \simeq \lambda_{\rm rad}^{-1})$,  as expected. From \eqref{eq:Eb} we can read off the RG flow, after time averaging, to find
\beq
\label{eq:Mrendef0}
\mu \frac{d}{d\mu} \langle E\rangle =0 \qquad \Longrightarrow \qquad \mu \frac{d}{d\mu} \left \langle M_{\rm ren}(t,\mu)\right\rangle = -  2G_N^2 M \left\langle I^{(3)}_{ij}(t) I^{(3)}_{ij}(t)\right\rangle\,.
\eeq
This expression is in agreement with the result in \cite{andirad3}, and leads to
\beq
\label{eq:Mrendef}
 \langle E \rangle = \left\langle M_{\rm ren}(t, \mu_0\simeq r^{-1})\right\rangle -  2G_N^2 M \left\langle I^{(3)}_{ij}(t) I^{(3)}_{ij}(t)\right\rangle\log v\,. 
\eeq
The 4PN logarithmic correction in the last term was first discussed in \cite{ALTlogx}.\vskip 4pt
 
Gathering all the pieces, we finally arrive at a balanace equation of the sort,
\beq
\langle \dot E (t) \rangle =  -P_{\rm local} -\frac{2 G_N^2 M}{5} \,\left \langle I^{ij(1)}(t)~{\rm PV}\, \int_{-\infty}^\infty dt' \,  I^{ij(6)}(t') \left[\frac{1}{t-t'}\right]\right\rangle\,,
\eeq
including the non-conservative part of the tail from \eqref{eq:tailanc}. Here, $P_{\rm local}$ represents the power loss induced by all other (local) dissipative terms (e.g., from Sec.~\ref{sec:rr}). Performing the time average using the leading expression for the quadrupole moment on a circular orbit with frequency $\Omega$, we recover the leading contribution to the power loss due to the tail effect, e.g.  \cite{andirad},
\beq
\label{eq:powertail}
\frac{P_{\rm tail}}{P_{LO}} = 4\pi x^{3/2}\,.
\eeq
Here $x \equiv (G_N M \Omega)^{2/3}$ is the standard PN expansion parameter. Notice the relevant factors of~$\pi$, which now appear through the study of radiation-reaction effects, but without the associated IR divergences discussed in \cite{andirad}. See next for more on this issue.

\section{Discussion}\label{sec:disc}

In this paper we computed the tail contribution to the gravitational radiation reaction to 4PN order within the EFT framework. We arrived at an effective action that displays time non-locality, i.e.~\eqref{finitetail}, contribuiting conservative as well as dissipative terms to the radiation-reaction force, i.e.~\eqref{eq:tailac} and \eqref{eq:tailanc}, respectively. The former being responsible for non-trivial RG trajectories for the gravitational binding potential and mass/energy in the near zone, i.e.~\eqref{eqVtm0} and \eqref{eq:Mrendef0}, while the latter leads to the well-known power loss due to the leading tail effect, i.e. \eqref{eq:powertail}.\vskip 4pt

Given the nature of the computation, naively, one would have thought that the tail contribution to the effective action could have been obtained by replacing the source quadrupole moment in the Burke-Thorne result \eqref{eq:arad}, with the corresponding {\it radiative} moment induced by the tail effect, e.g. \cite{andirad,srad}. However, while the leading tail contribution to the radiative quadrupole presents an IR divergence \cite{andirad,srad}, we find here instead a UV singularity, i.e.~\eqref{eq:RRnl}. As we argued, the singular behavior in the tail contribution to the effective action stems off the conservative~sector. The UV pole is thus ultimately canceled by a counter-term in the near zone, but arising from an IR divergence~\cite{4pn1,4pn2,4pndim}. This is consistent with the expectation that the counter-term must originate in the potential region, and moreover, that UV divergences in the near zone are renormalized through counter-terms  arising from the point-particle worldline action for the constituents of the binary. In both cases (radiative multipoles and radiation-reaction effects) the divergence is due to a $1/r$ long-range force. However, the fact that we are computing the tail contribution to the radiation-reaction force in an EFT where we treat the binary as a point-like object, transforms the expected IR into a UV behavior. In other words, a would-be logarithmic IR divergence, $\propto \log r$, is converted into a UV singularity, when $r \to 0$. This demonstrates one of the remarkable features of the EFT formalism, which allowed us to use the RG machinery to resum logarithms.\vskip 4pt Let us emphasize that there are no poles in the full theory calculation, which displays instead a logarithm of the ratio of physical scales. The divergences arise in the EFT side because of the separation into regions and the point-particle limit in \eqref{eq:lag}. The IR/UV poles cancel out, as~expected.  However, because of the introduction of an IR regulator, the arbitrariness of the different schemes leaves the result depending on an extra constant, $C$, at 4PN order \cite{4pn2,binigsf}. In~spite of this, the RG equations and long-distance logarithms are universal, and do not depend on the details of the matching at the orbit scale, i.e.~\eqref{eq:nontrV}. This analysis thus explains the origin of the logarithmic term found in \cite{ALTlogx,andirad3}, i.e.~\eqref{eq:Mrendef}.\vskip 4pt The reader may be puzzled about the appearance of this extra parameter, $C$. In principle, one should be able to compute the 4PN potential without the need of additional information. In~fact, the existence of IR divergences in the computation of the static potential is also known to occur in QCD, the theory of the strong interaction, amusingly called ADM singularities \cite{ADM} (after Appelquist, Dine and Muzinich). These singularities  re-appear in the EFT approach NRQCD, for non-relativistic quarks, and in particular when performing a matching computation into pNRQCD, where the potential is treated as a Wilson coefficient, somewhat similar to a multipole expansion, e.g. \cite{bramb1,bramb2,Hoang}. In this case, the IR divergences cancel out in the matching, without requiring extra conditions. The cancelation is due to contributions from two --in principle different-- regions, namely potential and {\it ultrasoft} modes. In other words, the IR behavior of the potentials, when $\bk \to 0$, overlaps with the contribution from softer modes, which in pNRQCD becomes a self-energy diagram as in Fig.~\ref{nltail}   (but with a propagating heavy field and without the~tail).\vskip 4pt 

These manipulations, translated into the classical limit, are strikingly similar to what we encounter here in NRGR, in particular for the matching of the binding potential. Moreover, we also find that the IR singularity in the near region cancels out against a pole in the radiation theory with long-wavelength fields, but instead of a UV nature. This is the reason why, in principle, different IR and UV regulators may introduce arbitrariness. However, as in QCD, the existence of these overlapping divergences is due to double-counting in the EFT.  This issue is ultimately related to the so called `zero-bin subtraction'\cite{zerobin}, which will be required in the ongoing computation of the 4PN potential \cite{nrgr4pn}. Once the double-counting is properly removed the static potential becomes an IR-safe quantity, and the necessity of additional information beyond the PN framework, advocated in \cite{4pnDam2}, disappears. The parameter $C$ will be then fixed by the left over finite pieces after the subtraction of the zero-bin.  As we emphasized, the long-distance logarithms and RG flow discussed here are not affected by this procedure. See \cite{zerobinNRGR} for more details.\vskip 4pt

Finally, unlike the computations in \cite{andirad3}, where five diagrams are required for the one-point function, the results obtained here --at the level of the effective action-- are derived from a single one, i.e.~Fig.~\ref{nltail}, and without the need of a four-graviton vertex. That is because computing the one-point function corresponds to attaching an external leg to the diagram in Fig.~\ref{nltail}, and there are five different ways to do so. Namely, two from attaching a leg to a propagator either sourced by $M$ or the quadrupole, two more from attaching a leg at the associated vertices, and another one from a four-graviton coupling \cite{andirad3}. This suggests that the analysis presented here may be more suitable to compute higher order contributions from tail effects and the resulting logarithmic corrections. We leave this possibility open for future work.

\subsubsection*{Relation to previous work}
 
 The computation of the tail effect within NRGR was first investigated in \cite{tailFoffa}, where an expression equivalent to \eqref{finitetail} was presented, see Eq.~(11) in \cite{tailFoffa}. However, several aspects of the calculations in \cite{tailFoffa}, and subsequent interpretation, are unfortunately either inconsistent or unjustified, which in part motivated us to write the present paper.  
For example, in Eq.~(12) of \cite{tailFoffa} we find an expression similar to ours in \eqref{eq:RRnl}. However, while we emphasized the term proportional to $i \pi \text{sign}(\omega)$, only a factor of $i\pi$ is written in  Eq.~(12) of \cite{tailFoffa}. This is inconsistent with the result quoted in Eq.~(11), nor does it properly incorporate the dissipative contribution from the tail effect. The computation in \cite{tailFoffa} is repeated in coordinate space in an appendix, resulting in the correct expression reported in Eq.~(11). Hence, we do not insist on this point as the main discrepancy between the authors' approach and ours. The main difference turns out to be the renormalization procedure.\vskip 4pt While we argue that the radiation-reaction force in the near zone is renormalized through a counter-term that originates as an IR singularity in the potential region \cite{4pn1}, instead in \cite{tailFoffa} a counter-term was written, $M_{\rm ct}$, for the binding mass/energy term in the effective action for the radiation theory, see their Eq.~(13). After introducing $M_{\rm ct}$ the effective action becomes finite in the $\epsilon \to 0$ limit, but at the same time one is forced --by imposing the $\mu$-independence of the effective action shown in their Eq.~(14)-- to write an RG equation for $M_{\rm ren}(\mu)$ (similarly to what we did in \eqref{eq:Vrenmu} for $V_{\rm ren}(\mu)$). The resulting RG flow for $M_{\rm ren}(\mu)$ would be incorrect, and disagrees with their own Eq.~(19), which is the one in agreement with our \eqref{eq:Mrendef0} and the result in \cite{andirad3}. Even ignoring this internal inconsistency, other manipulations are rather dubious. For example, the existence of an arbitrary extra parameter $\lambda$ (beyond the existing of the $\mu$ scale) in the expression for the binding energy in Eq.~(18) of \cite{tailFoffa}. The meaning of $\lambda$ is not apparent to us nor how its value is supposed to be fixed, especially given the claim ``for any $\lambda$'' \cite{tailFoffa} after its appearance in their Eq.~(16). This makes their reproduction of the logarithmic term at 4PN found in \cite{ALTlogx}, quoted in Eq.~(22) of \cite{tailFoffa}, unclear.\vskip 4pt 

In summary, we believe our work in this present paper clarifies and makes consistent how the divergences must be handled within NRGR, and how to systematically incorporate logarithmic corrections to the binding mass/energy.

 \section*{Acknowledgments}
 We thank Ira Rothstein for very helpful discussions. C.R.G. is supported by NSF grant PHY-1404569 to the California Institute of Technology and also thanks the Brinson Foundation for partial support. A.K.L. is supported by NSF grant PHY-1519175. R.A.P is supported by the Simons Foundation and S\~ao Paulo Research Foundation (FAPESP) Young Investigator Awards, grants 2014/25212-3 and 2014/10748-5. A.R. was supported by NASA grant 22645.1.1110173.

\appendix

\section{Calculation of the 4PN tail contribution to radiation-reaction}\label{sec:app1}

The calculation for the tail effect arises from a diagram with a mass insertion, a triple-graviton vertex, and three propagators, as shown in~Fig.~\ref{nltail}. The latter comes in $2^4 = 16$ different combinations of history indices (i.e., $\pm$) for each vertex.  The effective action is given by\begin{align}
 i W & = \int \frac{d \omega}{2 \pi} \int_{\bk, \bq} M\,I^{ij}_-(- \omega) I^{ij}_+(\omega) V_{\bar h\bar h\Phi}(\omega, \bk, \bq) \frac{i}{- \bq^2 } \frac{i}{(\omega + i \epsilon)^2- \bk^2} \frac{i}{(\omega + i \epsilon)^2- (\bk + \bq)^2}  \label{eq_Sefftail2}
\end{align}
in the $\pm$ doubled variables and we dropped higher order terms in the minus variables as they do not contribute to equations of motion \cite{chadprl}. Note that the 3-graviton vertex $V_{\bar h\bar h\Phi}$ does not depend on the history labels.\vskip 4pt  When we work in momentum space, it is essential to impose the correct momentum routing corresponding to the given retarded boundary conditions. If we follow the usual recipe and replace derivatives $\partial_\mu$ by $-i k_\mu$ where $k$ is the {\it incoming} 4-momentum, we find that the 4-momentum flows through a retarded propagator $G_{\text{ret}}(x-y)$ from the earlier event, $y$, to the later one, $x$. Moreover, wherever a momentum $(\omega, \bk)$ flows into a worldline vertex coupling to the quadrupole, the latter depends on the frequency as $I^{ij}(-\omega)$ whereas at a quadrupole vertex where a 4-momentum $(\omega, \bk)$ flows out, we have $I^{ij}(\omega)$. Finally, since the mass, $M$, can be taken to be time-independent up to higher orders, the propagator coupling to the mass is the usual static Newton-like term, i.e.~\eqref{Phist}.  These conventions result in the momentum routing shown~in~Fig.~\ref{nltail}.\vskip 4pt  
To include all proper momentum and tensor structures is rather messy and not very illuminating. The resulting expression takes the general form
\begin{align}
 i W & = \int \frac{d \omega}{2 \pi} \int_{\bk, \bq}  \frac{f(\omega, \bk, \bq)}{q^2 k^2 (k+q)^2} \, , \label{eq_Sefftail3}
\end{align}
where the four-vectors used in the denominator are $q^\mu = (0, \bq)$ and $k^\mu = (\omega, \bk)$, so with our convention for the metric we have \bea q^2 &=& - \bq^2\,, \\ k^2 &=& (\omega + i \epsilon)^2 - \bk^2\,\\ (k+q)^2 &=& (\omega + i \epsilon)^2 - (\bk + \bq)^2\,.
\eea
The function in the numerator of (\ref{eq_Sefftail3}), $f(\omega, \bk, \bq)$,  is proportional to $I^{ij}_-(- \omega) I^{lm}_+(\omega)$ and contains up to four momenta contracted with the four indices of the two quadrupole moments. It~also contains scalar products, all of which can be written in terms of squares, e.g.  $k \cdot q = \frac{1}{2}\left[(k+q)^2 - k^2 - q^2\right]$. Notice any factor of $q^2$, $k^2$ or $(k+q)^2$ in the numerator cancels against one of the propagators in the denominator. Moreover, except for $q^2$, the other two lead to a scale-less integral which can be set to zero in dim.~reg. Therefore, we can write the resulting expression as a sum of two terms, 
\begin{align}
 i W & = \int \frac{d \omega}{2 \pi} \int_{\bk, \bq}  \frac{f_3(\omega, \bk, \bq)}{q^2 k^2 (k+q)^2}  + \int \frac{d \omega}{2 \pi} \int_{\bk, \bq}  \frac{f_2( \omega, \bk, \bq)}{k^2 (k+q)^2}\,.
   \label{eq_Sefftail4} \
\end{align}
Beginning  with the piece with two factors in the denominator, we find
\begin{align}
& \hspace{0.8cm}\int \frac{d \omega}{2 \pi} \int_{\bk, \bq}  \frac{f_2( \omega, \bk, \bq)}{k^2 (k+q)^2}   =   \int \frac{d \omega}{2 \pi} \int_{\bk, \bq}  \frac{f_2( \omega, \bk, \bq-\bk)}{k^2 q^2} = \notag \\
&   \frac{i M}{32 m_{Pl}^4} \int \frac{d \omega}{2 \pi} \int_{\bk, \bq} \!\!\! I^{ij}_-(- \omega) I^{lm}_+(\omega) \frac{\omega^2 \delta^{il} \bk^j \bq^m - \frac{(d-3)^2}{(d-2)^2} \bk^i \bk^j \bq^l \bq^m}{[(\omega + i \epsilon)^2 - \bk^2] [(\omega + i \epsilon)^2 - \bq^2]}  \, . \label{eq_Sefftail3a} \
\end{align}
Note that there is no term proportional to $\delta^{il}\delta^{jm}$. The one with a single $\delta^{il}$ vanishes because the double integration factorizes into two pieces which are linear in $\bk$ and $\bq$, with no preferred direction. The last term in the numerator vanishes because the resulting integral traces over trace-free quadrupoles. We are thus left with the piece of the effective action in \eqref{eq_Sefftail4} with three propagators. The different possible integrals can be reduced as follows,
\begin{align}
 \int_{\bk, \bq} \frac{\bk^i \bk^j}{q^2 k^2 (k+q)^2} & = \frac{\omega^2 J_0}{d-1} \, \delta^{ij}, \\
 \int_{\bk, \bq} \frac{\bk^i \bq^j}{q^2 k^2 (k+q)^2} & = \frac{I_0}{2(d-1)} \, \delta^{ij},  \\
 \int_{\bk, \bq} \frac{\bq^i \bq^j}{q^2 k^2 (k+q)^2} & = - \frac{I_0}{d-1} \, \delta^{ij}, \\
 \int_{\bk, \bq} \frac{\bk^i \bk^j \bk^l \bk^m}{q^2 k^2 (k+q)^2} & = \frac{\omega^4 J_0}{(d-1)(d+1)} \left(\delta^{ij} \delta^{lm} + \delta^{il} \delta^{jm} + \delta^{im} \delta^{jl}\right), \\
 \int_{\bk, \bq} \frac{\bk^i \bk^j \bk^l \bq^m}{q^2 k^2 (k+q)^2} & = \frac{\omega^2 I_0}{2(d-1)(d+1)} \left(\delta^{ij} \delta^{lm} + \delta^{il} \delta^{jm} + \delta^{im} \delta^{jl}\right), \\
 \int_{\bk, \bq} \frac{\bk^i \bk^j \bq^l \bq^m}{q^2 k^2 (k+q)^2} & = - \frac{\omega^2 I_0}{(d-2)(d+1)} \, \delta^{ij} \delta^{lm} - \frac{(d-3) \omega^2 I_0}{2 (d-2)(d-1)(d+1)} \left( \delta^{il} \delta^{jm} + \delta^{im} \delta^{jl}\right),\\
 \int_{\bk, \bq} \frac{\bk^i \bq^j \bq^l \bq^m}{q^2 k^2 (k+q)^2} & = \frac{\omega^2 I_0}{(d-1)(d+1)} \left(\delta^{ij} \delta^{lm} + \delta^{il} \delta^{jm} + \delta^{im} \delta^{jl}\right), \\
 \int_{\bk, \bq} \frac{\bq^i \bq^j \bq^l \bq^m}{q^2 k^2 (k+q)^2} & = - \frac{2 \omega^2 I_0}{(d-1)(d+1)} \left(\delta^{ij} \delta^{lm} + \delta^{il} \delta^{jm} + \delta^{im} \delta^{jl}\right) .
\end{align} 
\vskip 4pt
\noindent where
\begin{align}
\label{eq:I0}
 I_0 & =   \int_{\bk, \bq} \frac{1}{[(\omega + i \epsilon)^2 - \bk^2] [(\omega + i \epsilon)^2 - (\bk + \bq)^2]}  \\
             &= \frac{\left(\Gamma \left[-\frac{d-3}{2}\right]\right)^2}{(4 \pi)^{d-1}} \left[-(\omega + i \epsilon)^2\right]^{d-3} \, ,\nn\\
 J_0 & = \int_{\bk, \bq} \frac{1}{[- \bq^2] [(\omega + i \epsilon)^2 - \bk^2] [(\omega + i \epsilon)^2 - (\bk + \bq)^2]}\label{eq:J0}\\ &= - \frac{1}{d-4} \, \frac{\Gamma \left(-\frac{d-3}{2}\right) \Gamma \left(-\frac{d-5}{2}\right)}{(4 \pi)^{d-1}} \left[-(\omega + i \epsilon)^2\right]^{d-4} \nn\,.
\end{align}
The effective action then reads,
\begin{align}
 i W = - i\int \frac{d\omega}{2 \pi} \,  \frac{ (d-3) M \omega^4 \, I_-^{ij}(-\omega) I_+^{ij}(\omega)}{32 (d-2)^2 (d-1)(d+1)} \left[(d^2 - 2 d + 3) I_0 + d(d-2)(d-1) \omega^2 J_0 \right]\,,
\end{align}
and expanding around $d = 4$, we arrive at the expression in \eqref{eq:RRnl}. 

\newpage
\bibliographystyle{utphys}
\bibliography{RefsNL}

\end{document}